\documentclass[preprint,12pt,3p]{elsarticle}

\usepackage{matlab-prettifier}
\usepackage{pdflscape}
\usepackage{graphicx}
\usepackage{makecell}
\usepackage{xfrac}
\usepackage{parskip}
\usepackage[backref=page]{hyperref}
\usepackage{adjustbox}
\usepackage{setspace}
\hypersetup{ 
backref=true, 
bookmarks=true, 
pdftoolbar=true, 
pdfmenubar=true, 
pdffitwindow=false, 
pdfstartview={FitH}, 
pdftitle={}, 
pdfauthor={}, 
pdfsubject={}, 
pdfkeywords={}{}, 
pdfnewwindow=true, 
colorlinks=true, 
linkcolor=BlueViolet, 
citecolor=maroon, 
urlcolor=blue,
}
\usepackage{comment}
\usepackage{subfig}
\usepackage{caption}
\usepackage{longtable}
\usepackage{amssymb}
\usepackage{amsmath}
\usepackage{gensymb}
\usepackage{siunitx}

\usepackage{natbib}
\biboptions{comma,round}
\setcitestyle{authoryear}

\begin{document}
\begin{frontmatter}

\title{Meta-emulation: An application to the social cost of carbon}

\author[label1,label2,label3,label4,label5,label6]{Richard S.J. Tol\corref{cor1}
}
\address[label1]{Department of Economics, University of Sussex, Falmer, United Kingdom}
\address[label2]{Institute for Environmental Studies, Vrije Universiteit, Amsterdam, The Netherlands}
\address[label3]{Department of Spatial Economics, Vrije Universiteit, Amsterdam, The Netherlands}
\address[label4]{Tinbergen Institute, Amsterdam, The Netherlands}
\address[label5]{CESifo, Munich, Germany}
\address[label6]{Payne Institute for Public Policy, Colorado School of Mines, Golden, CO, USA}

\cortext[cor1]{Jubilee Building, BN1 9SL, UK}

\ead{r.tol@sussex.ac.uk}
\ead[url]{http://www.ae-info.org/ae/Member/Tol\_Richard}

\begin{abstract}
A large database of published model results is used to estimate the distribution of the social cost of carbon as a function of the underlying assumptions. The literature on the social cost of carbon deviates in its assumptions from the literatures on the impacts of climate change, discounting, and risk aversion. The proposed meta-emulator corrects this. The social cost of carbon is higher than reported in the literature.
\textit{Keywords}: social cost of carbon; meta-analysis; quantile regression\\
\medskip\textit{JEL codes}: Q54
\end{abstract}

\end{frontmatter}

\section{Introduction}
The literature on the social cost of carbon is large, with myriad assumptions, many contested. Policy advice on the social cost of carbon typically relies on a single model \citep{Stern2006} or a handful \citep{IAWGSCC2021} instead of reflecting the breadth and depth of scholarship. I here propose a method\textemdash meta-emulation\textemdash that respects the entire body of research on the social cost of carbon but allows policymakers to impose their preferred parameter choices. I illustrate the proposed method with alternative estimates for the discount rate and the impact of climate change.

Meta-analysis is a versatile method to make inferences about a body of literature. In a previous paper \citep{Tol2023NCC}, I used meta-analysis to estimate the distribution of published estimates of the social cost of carbon, showing that the distribution has shifted statistically significantly upwards over time. In another recent paper \citep{Tol2025anyas}, I found evidence of selective citation and clustering on particular assumptions. More generally, meta-analysis reveals the relationship between the variable of interest and the methods and assumptions used to estimate its value. Meta-analysis can be used to test the hypothesis that certain results are less likely to be published \citep{Havranek2015}, but it cannot reveal whether the underlying assumptions are justified or representative. However, it can reveal, through the method of meta-emulation proposed here, how outcomes change with different assumptions and test whether the difference is statistically significant.

\citet{Moore2024PNAS} have goals similar to this paper, but data, methods, and intentions differ. Moore use 1,823 estimates from 147 papers. I use 14,152 estimates from 446 papers. They excluded many papers because they were too old or too new. They included only journal articles,\footnote{They also included a handful of papers that do not estimate the social cost of carbon but report its shadow price instead.} even though a lot of relevant material is published in books and reports, the latter being particularly influential on policy. They used random forests which, if applied judiciously,\footnote{Some of their cell sizes are too small for my taste.} work well in-sample but are less suited for extrapolation. I use regression instead for better out-of-sample performance. They focus on the central tendency whereas I consider the entire distribution of the social cost of carbon. Finally, while both Moore and I show sensitivity to key assumptions but I do not dictate what those assumptions should be.\footnote{For example, Moore argues in favour of stochastic models. However, \citet{EstradaTol2015, Estrada2025anyas} show that these models tend to be miscalibrated.} I instead borrow alternative assumptions from the relevant literature.

The paper proceeds as follows. Data and methods are discussed in Section \ref{sc:data}, results in Section \ref{sc:data}. Section \ref{sc:conclude} concludes.

\section{Data and methods}
\label{sc:data}

\subsection{Data}
\citet{Tol2025data} records 14,152 estimates of the social cost of carbon, published in 446 papers. The database has many fields but I here use only five: The social cost of carbon for 2025 emissions in 2024 US dollars per metric tonne of carbon; the year of publication; the pure rate of time preference; the inverse of the elasticity of intertemporal substitution; and the impact of 2.5\celsius{} warming. Following \citep{Tol2023NCC}, papers are quality-weighted; within papers, estimates that are favoured by the authors are emphasized; implausible estimates are discounted, inconsistent ones disregarded.

Figure \ref{fig:scc} shows the distribution of estimates of the social cost of carbon. Some 1\% of estimates point to a benefit of greenhouse gas emissions. The mode lies between \$75 and \$100/tC, the mean around \$400/tC. The distribution is right-skewed, with a thick, perhaps fat tail \citep{Anthoff2022}.

\citet{Drupp2022} and \citet{Nesje2023} survey experts on the appropriate time preferences for public policy. \citet{Havranek2015JIE} meta-analyzes published estimates of the elasticity of intertemporal substitution. \citet{Matousek2022} do the same for estimates of the discount rate, focusing on experiments with short time horizons; as income growth is minimal, the discount rate can be interpreted as the pure rate of time preference. \citet{Tol2024EnPol} surveys estimates of the economic impact of climate change. The data for all five studies are in the public domain.

Figure \ref{fig:timehist} compares and contrasts the assumptions made in the literature on the social cost of carbon on the pure rate of time preference to the opinions of experts, as surveyed by \citet{Drupp2022} and \citet{Nesje2023}, and the literature, as surveyed by \citet{Havranek2015JIE}. The distributions differ. The social cost of carbon literature shows bunching around rates of 0\%, 1\%, 1.5\%, and 3\%. The other distributions are more spread out, in different ways. Matousek and Nesje find lower values than Drupp, while Nesje finds more support for intermediate values.

Figure \ref{fig:riskhist} repeats Figure \ref{fig:timehist} for the inverse of the elasticity of intertemporal substitution, using \citet{Drupp2022}, \citet{Nesje2023} and \citet{Matousek2022}. The distributions are again different. The social cost of carbon has a pronounced model around 1.5, whereas the experts by Drupp and Nesje prefer 1. Havranek's distribution is fairly flat, which a mode at a linear utility function.

Figure \ref{fig:imphist} does the same for the economic impact of climate change as surveyed in \citet{Tol2024EnPol}. The central estimate in the social cost of carbon literature is too pessimistic, while it ignores both positive and negative outliers in the economic impact literature.

All three figures reveal that the assumptions made in the literature on the social cost of carbon are at odds with the revealed wisdom in the literature on its key assumptions. In this paper, I do not argue who is right or wrong. I use the difference to illustrate the proposed method.

\subsection{Methods}
\begin{figure}
    \centering
    \caption{The histogram of published estimates of the social cost of carbon.}
    \label{fig:scc}
    \includegraphics[width=\textwidth]{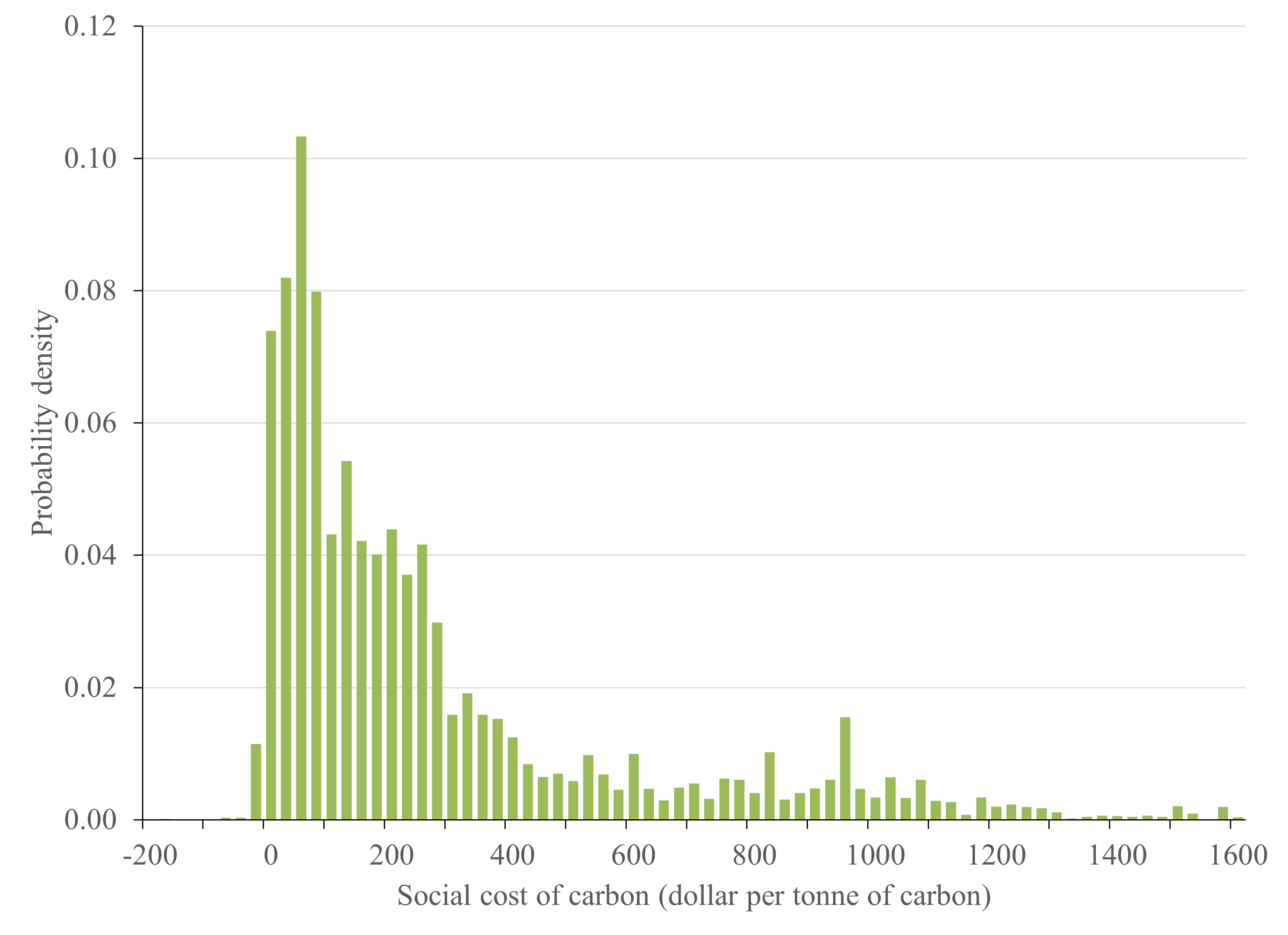}
\end{figure}

\begin{figure}
    \centering
    \caption{The histograms of the pure rate of time preference as used in the literature on the social cost of carbon, as preferred by scholars, and as measured.}
    \label{fig:timehist}
    \includegraphics[width=\textwidth]{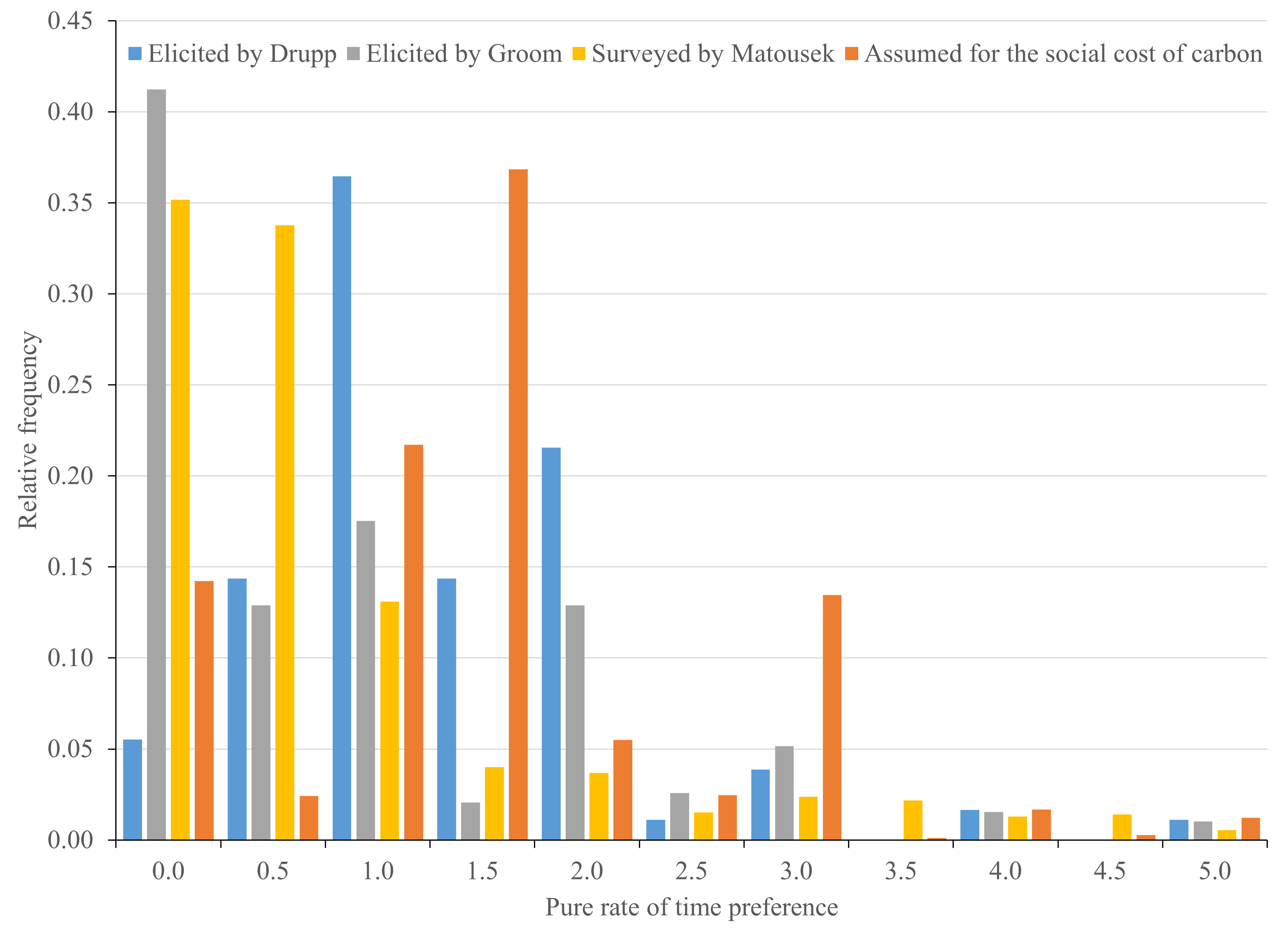}
\end{figure}

\begin{figure}
    \centering
    \caption{The histograms of the elasticity of the marginal utility of consumption as used in the literature on the social cost of carbon, as preferred by scholars, and as measured.}
    \label{fig:riskhist}
    \includegraphics[width=\textwidth]{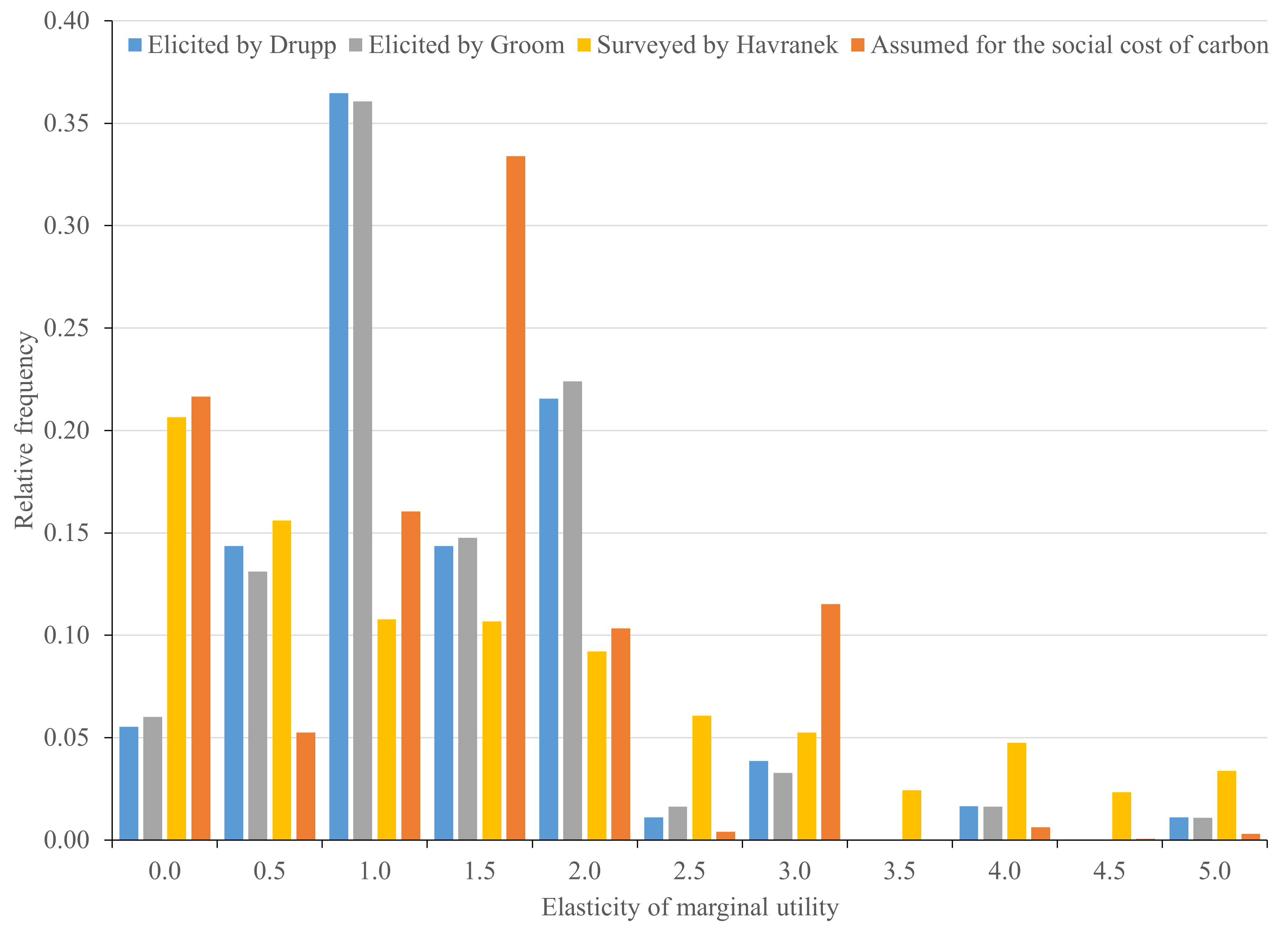}
\end{figure}

\begin{figure}
    \centering
    \caption{The histograms of the economic impact as used in the literature on the social cost of carbon and found in the literature on the impact of climate change.}
    \label{fig:imphist}
    \includegraphics[width=\textwidth]{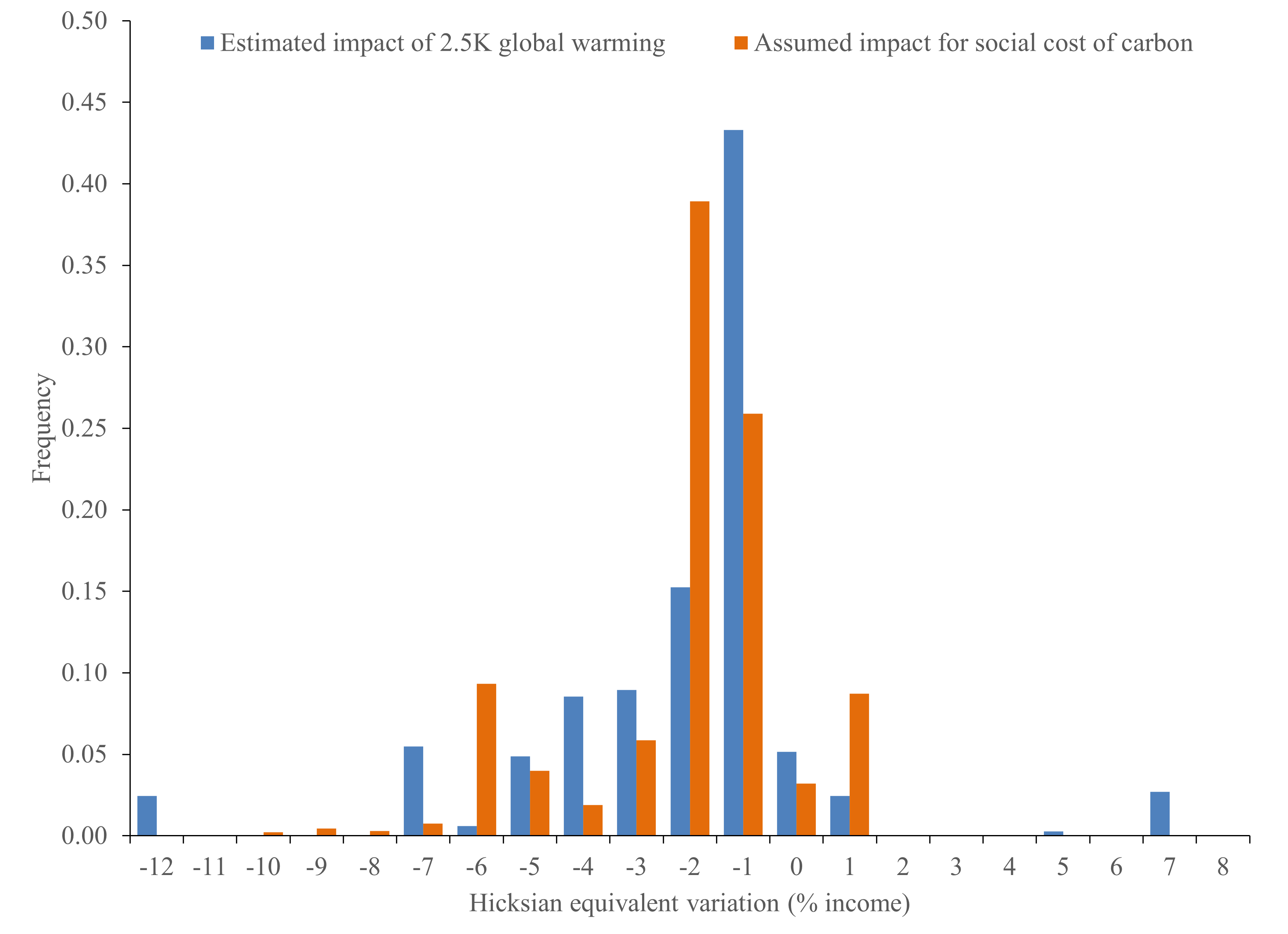}
\end{figure}

I use quantile regression to estimate the sensitivity of the distribution of the social cost of carbon to three key assumptions: the pure rate of time preference, the inverse of the elasticity of intertemporal substitution, and the impact of 2.5\celsius{} warming, controlling for the year of publication. Regressions are censored and weighted as in \citet{Tol2023NCC}. Quantile regression reveals how the entire distribution of the social cost of carbon changes. Figure \ref{fig:scc} shows the pronounced right tail. Limiting the attention to the central tendency would miss a key aspect of the concern about climate change. 

I then use the meta-analytic function to emulate what the distribution of the social cost of carbon would have been, had different assumptions been used. Specifically, 
\begin{equation}
    SCC_{a,p} = SCC_p + \sum_s \left ( F_{a,s} - P_{a,s} \right ) X_{a,s} \beta_p
\end{equation}
where $SCC_{a,p}$ is the \emph{predicted} social cost of carbon for altered assumption $a$, $SCC_p$ is the \emph{observed} social cost of carbon in percentile $p$, $\beta_p$ is the vector of estimated coefficients for that percentile, $X_{a,s}$ is the vector of possible values $s$ for assumption $a$, $F_{a,s}$ is the observed frequency for assumption $a$, and $P_{a,s}$ is the frequency of $s$ according to the expert elicitation or meta-analysis.

The standard error of the difference between two alternative views on the state of the world $P_{a,s}$ and $F_{a,s}$ for a \emph{single} assumption $a$ follows from
\begin{equation}
    \mathbb{V}\text{ar} ( SCC_{a,p} - SCC_{p} ) = \sum_s (F_{a,s} - P_{a,s})^2 \mathbb{V}\text{ar} X_{a,s} \beta_{a,p} = 
    \sigma_{a,p}^2 \sum_s (F_{a,s} - P_{a,s})^2 X_{a,s}^2 
\end{equation}
The standard error for the difference in views on \emph{multiple} assumptions would account for the correlations between estimates $F_{a1,s}$ and $F_{a2,s}$, between views $P_{a1,s}$ and $P_{a2,s}$, and between parameters $\beta_{a1,p}$ $\beta_{a2,p}$.

\section{Results}
\label{sc:results}

\subsection{The meta-emulator}
\begin{figure}
    \centering
    \caption{The impact of the assumed pure rate of time preference on the distribution of published estimates of the social cost of carbon.}
    \label{fig:prtp}
    \includegraphics[width=\textwidth]{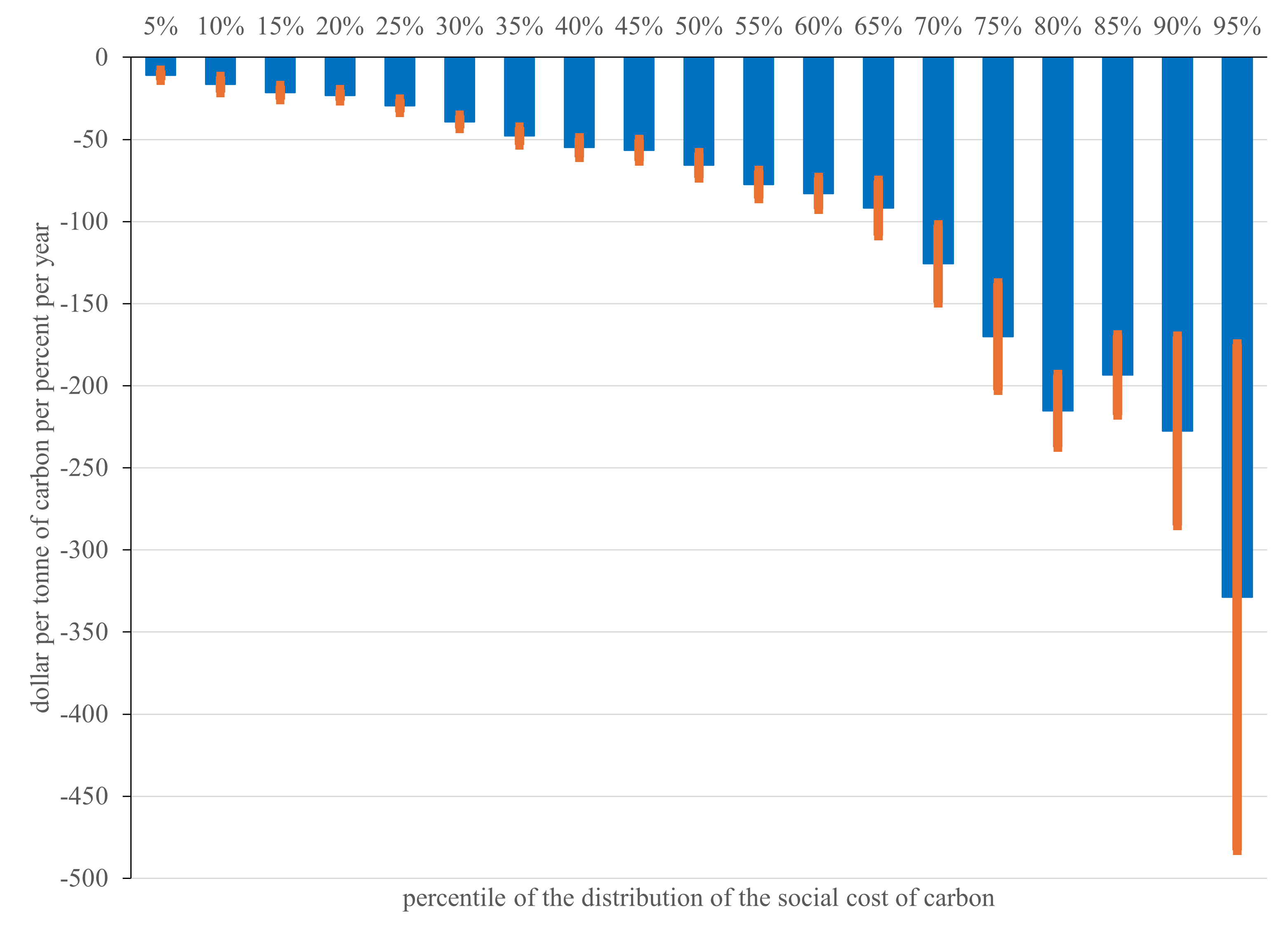}
\end{figure}

Figure \ref{fig:prtp} shows the results of the quantile regression of the social cost of carbon on the pure rate of time preference. Table \ref{tab:coeffs} has the numerical results. Greater impatience implies a lower social cost of carbon\textemdash this has been previously documented many times and follows immediately from first principles. At the median, an increase of the pure rate of time preference by 1\% leads to a drop in the social cost of carbon of \$66/tC. This is \$11/tC at the 5th \%ile, and \$329/tC at the 95th \%ile. The right tail fattens as the discount rate falls because the further future is more uncertain. 

\begin{figure}
    \centering
    \caption{The impact of the inverse of the assumed elasticity of intertemporal substitution on the distribution of published estimates of the social cost of carbon.}
    \label{fig:eis}
    \includegraphics[width=\textwidth]{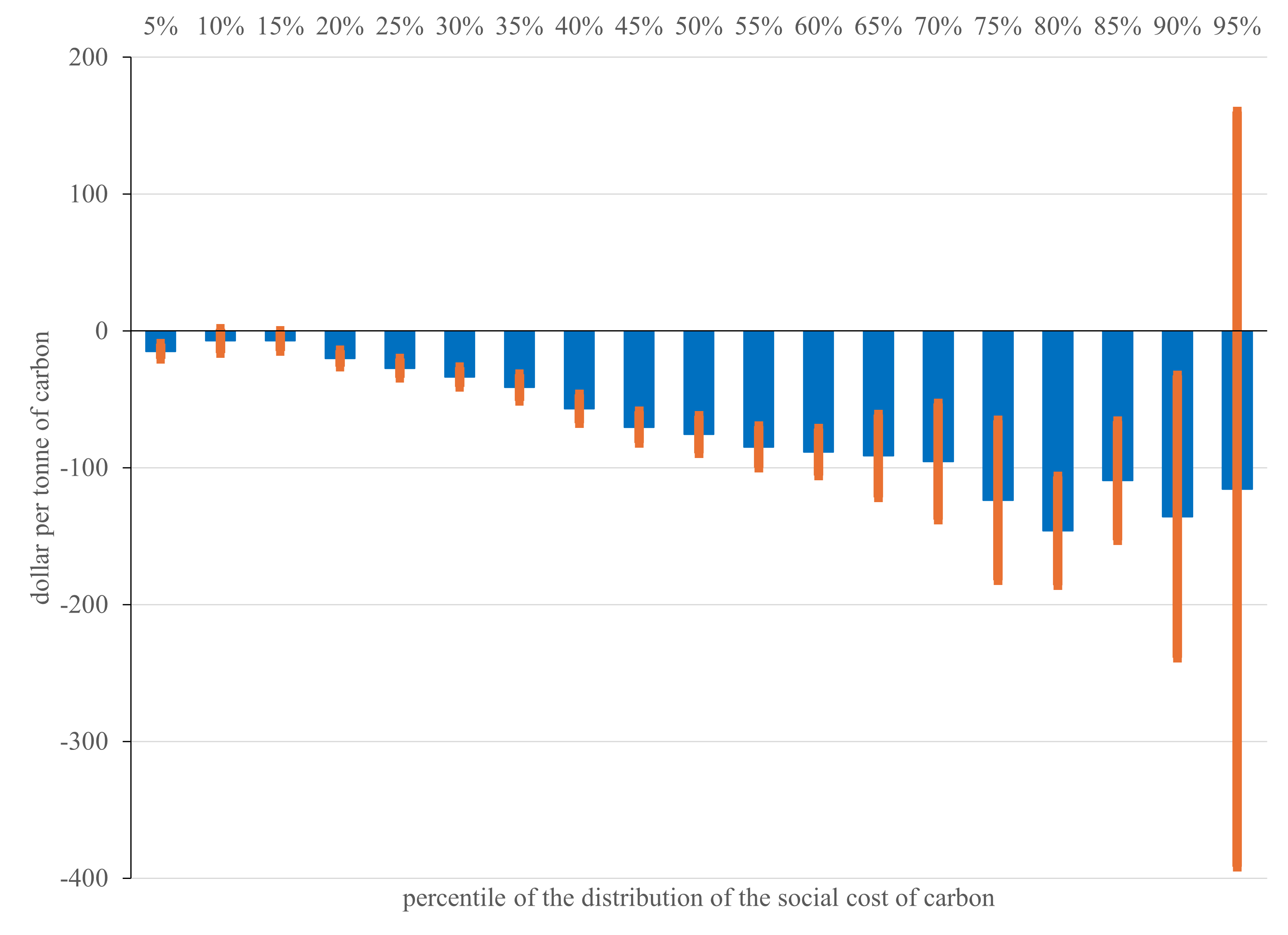}
\end{figure}

Figure \ref{fig:eis} shows the estimated parameters for the inverse of the elasticity of intertemporal substitution. See also Table \ref{tab:coeffs}. A stronger curvature of the utility function is associated with a lower social cost of carbon. An increase by 1 reduces the social cost of carbon by \$76/tC at the median, by \$20/tC at the 20th \%ile, and by \$109/tC at the 85th \%ile. The coefficients are not statistically significant at the 5\% level further in the tails.

\begin{figure}
    \centering
    \caption{The impact of the assumed economic impact of climate change on the distribution of published estimates of the social cost of carbon.}
    \label{fig:benchmark}
    \includegraphics[width=\textwidth]{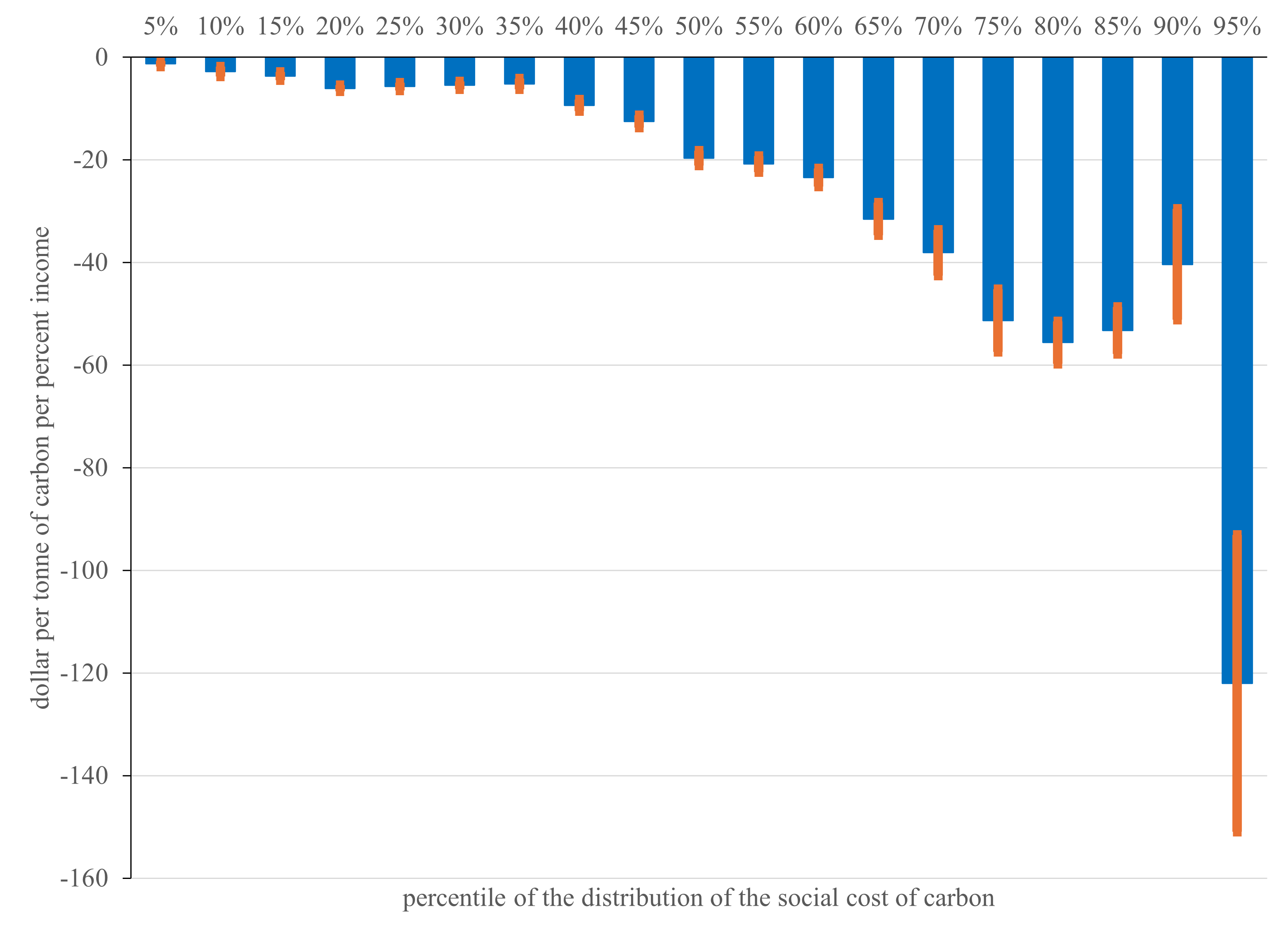}
\end{figure}

Figure \ref{fig:benchmark} shows the effects of the assumed economic impact at 2.5\celsius{} global warming. See Table \ref{tab:coeffs}. If the impact is 1\% of GDP worse, the median of the social cost of carbon increases by \$20/tC. This effect falls to \$1/tC at the 5th \%ile, and rises to \$122/tC at the 95th \%ile.

All three figures confirm that parameter variations change the \emph{distribution} of the social cost of carbon. Particularly, the right tail is more sensitive than the left tail. Risk-averse decision-makers would be concerned about this.

\subsection{Individual alternatives}
\begin{figure}
    \centering
    \caption{Cumulative distributions of the social cost of carbon following the literature and Drupp's experts on the pure rate of time preference.}
    \label{fig:emultime}
    \includegraphics[width=\textwidth]{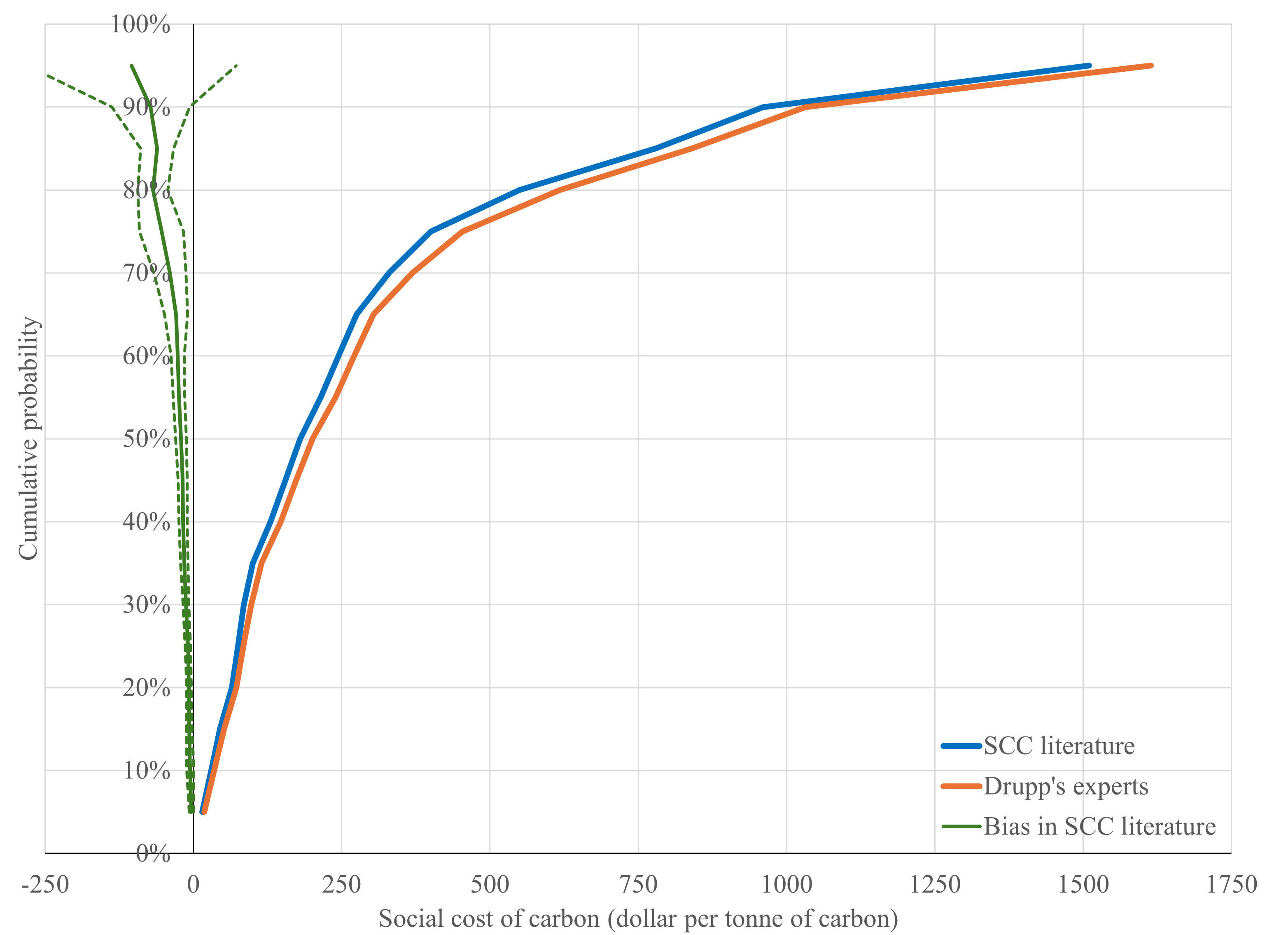}
    \caption*{\scriptsize The figure shows the empirical cumulative distribution of the social cost of carbon and the emulated cumulative distribution with the recommendations on time preference of Drupp's experts. The difference between the two is also shown with its 95\% confidence interval.}
\end{figure}

Figure \ref{fig:emultime} illustrates the proposed algorithm for the pure rate of time preference and Drupp's experts. The empirical cumulative distribution function of the social cost of carbon is very close to the emulated CDF following the recommendations by the survey. However, because of the perfect correlation between the other assumptions that drive the social cost of carbon, the difference due to the pure rate of time preference is nonetheless statistically significant. At the median, the literature underestimates the social cost of carbon by \$21/tC. This goes down to \$3/tC at the 5th \%ile, up to \$104/tC at the 95th \%ile.

\begin{figure}
    \centering
    \caption{The difference between the empirical and emulated cumulative distributions of the social cost of carbon due to the pure rate of time preference.}
    \label{fig:emultime2}
    \includegraphics[width=\textwidth]{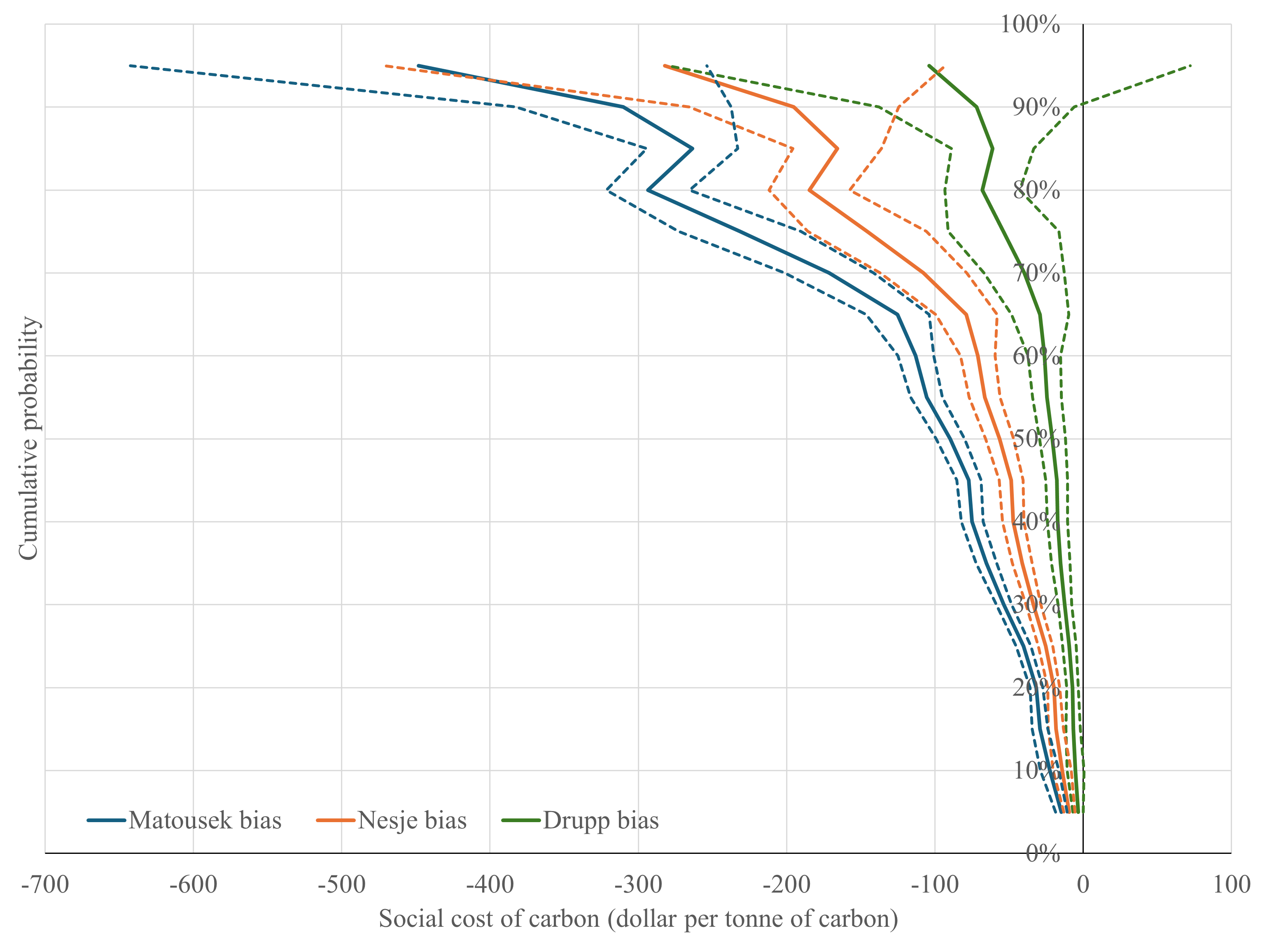}
    \caption*{\scriptsize The figures show the difference between the empirical cumulative distribution of the social cost of carbon and the emulated cumulative distribution with the recommendations on time preference of Drupp's experts and Nesje's experts and the revealed preferences in Matousek's meta-analysis. The dotted lines denote the 95\% confidence intervals.}
\end{figure}

Figure \ref{fig:emultime2} repeats the difference for Drupp's experts and adds Nesje's experts\textemdash the difference is larger, \$56/tC compared to Drupp's \$21/tC at the median\textemdash and Matousek's meta-analysis\textemdash the difference is larger still, \$90/tC at the median. Compared to these three alternatives, the literature on the social cost of carbon assumes too much impatience.

It is hard to choose between the three alternatives. Drupp is the most conservative, but based on stated preferences of a non-representative sample. Nesje's sample is smaller and arguably even less representative of the general population. Matousek uses revealed preferences and a much larger sample but is about private discounting in the short run\textemdash Drupp and Nesje are about public discounting in the long run. Drupp, Nesje and Matousek agree that the social cost of carbon literature is downward biased, but disagree by how much.

\begin{figure}
    \centering
    \caption{The difference between the empirical and emulated cumulative distributions of the social cost of carbon due to the elasticity of intertemporal substitution.}
    \includegraphics[width=\textwidth]{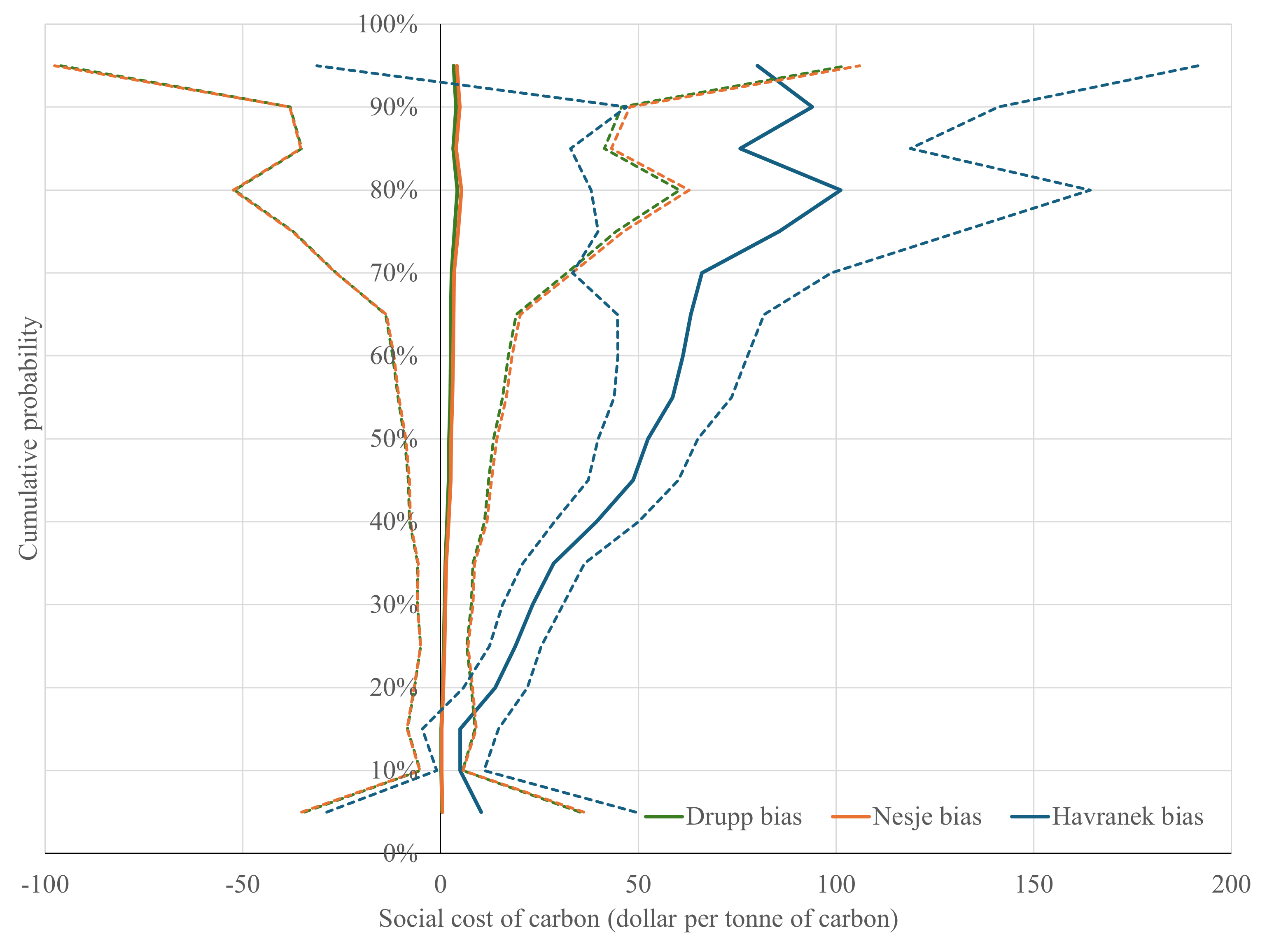}
    \caption*{\scriptsize The figures show the difference between the empirical cumulative distribution of the social cost of carbon and the emulated cumulative distribution with the recommendations on the inverse of the elasticity of intertemporal substitution of Drupp's experts and Nesje's experts and the revealed preferences in Havranek's meta-analysis. The dotted lines denote the 95\% confidence intervals.}
    \label{fig:emulrisk}
\end{figure}

Figure \ref{fig:emulrisk} shows the impact of alternative assumptions on the inverse of the elasticity of intertemporal substitution. The experts elicited by Drupp and Nesje largely agree with each other and with the assumptions in the social cost of carbon literature. Differences are insignificant. Havranek's meta-analysis, on the other hand, points to higher values of this inverse and hence to a lower social cost of carbon. The social cost of carbon in the literature is \$52/tC too high at the median, \$14/tC at the 20th \%ile, and \$94/tC at the 90th \%ile.

Assumptions about the economic impact of 2.5\celsius{} global warming also make a statistically significant difference, at least in the middle of the distribution of the social cost of carbon, as shown in Figure \ref{fig:emulimp}. Specifically, the literature on the social cost of carbon is biased towards optimistic assumptions on the total impact of climate change and thus underestimates the social cost of carbon.

\subsection{Joint alternatives}
Above, I vary one assumption at a time. Figure \ref{fig:twoway} shows the impact of alternative assumptions on the pure rate of time preference and the inverse of the elasticity of intertemporal substitution. This is not a simple addition. First, these two parameters correlate in the social cost of carbon literature. This explains why the sum of the Havranek and Matousek bias (+\$52/tC - \$45/tC = \$8/tC at the medium) is larger than the combined bias (\$5/tC). Second, the parameters correlate in the expert surveys too. The sum of the Drupp (Nesje) biases is -\$8/tC (-\$25/tC) while the combined bias is -\$29/tC (-\$49/tC).

\begin{figure}
    \centering
    \caption{The marginal impact of assumptions on the benchmark impact of climate change on the cumulative distribution of the social cost of carbon.}
    \label{fig:emulimp}
    \includegraphics[width=\textwidth]{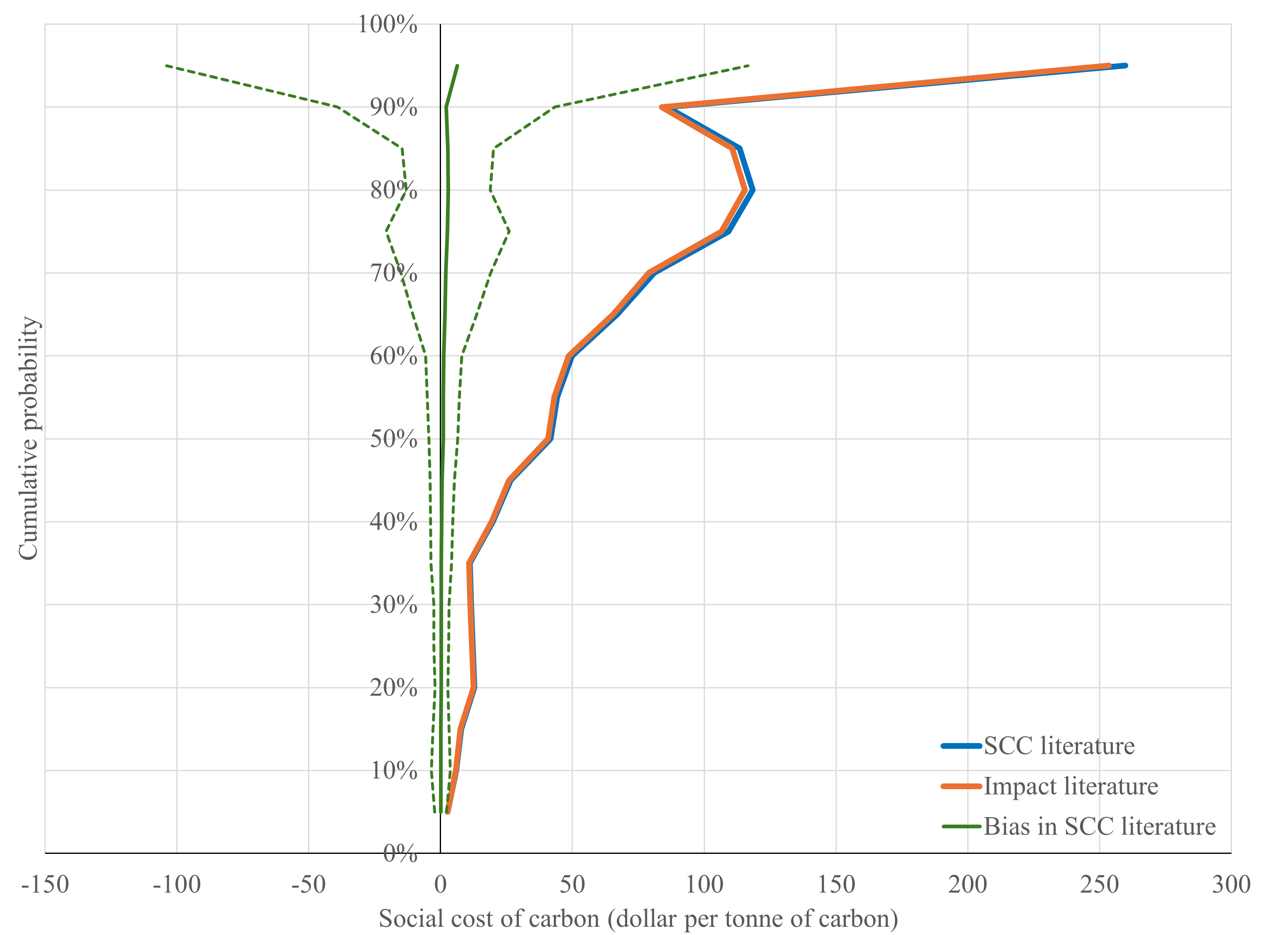}
    \caption*{\scriptsize The figure shows the marginal impact of the economic impact of 2.5\celsius warming on the cumulative distribution of the social cost of carbon for the assumptions in the social carbon literature and the economic impact literature. The difference between the two is also shown with its 95\% confidence interval.}
\end{figure}

\begin{figure}
    \centering
    \caption{The marginal impact of assumptions on the parameters of the Ramsey rule on the cumulative distribution of the social cost of carbon.}
    \label{fig:twoway}
    \includegraphics[width=\textwidth]{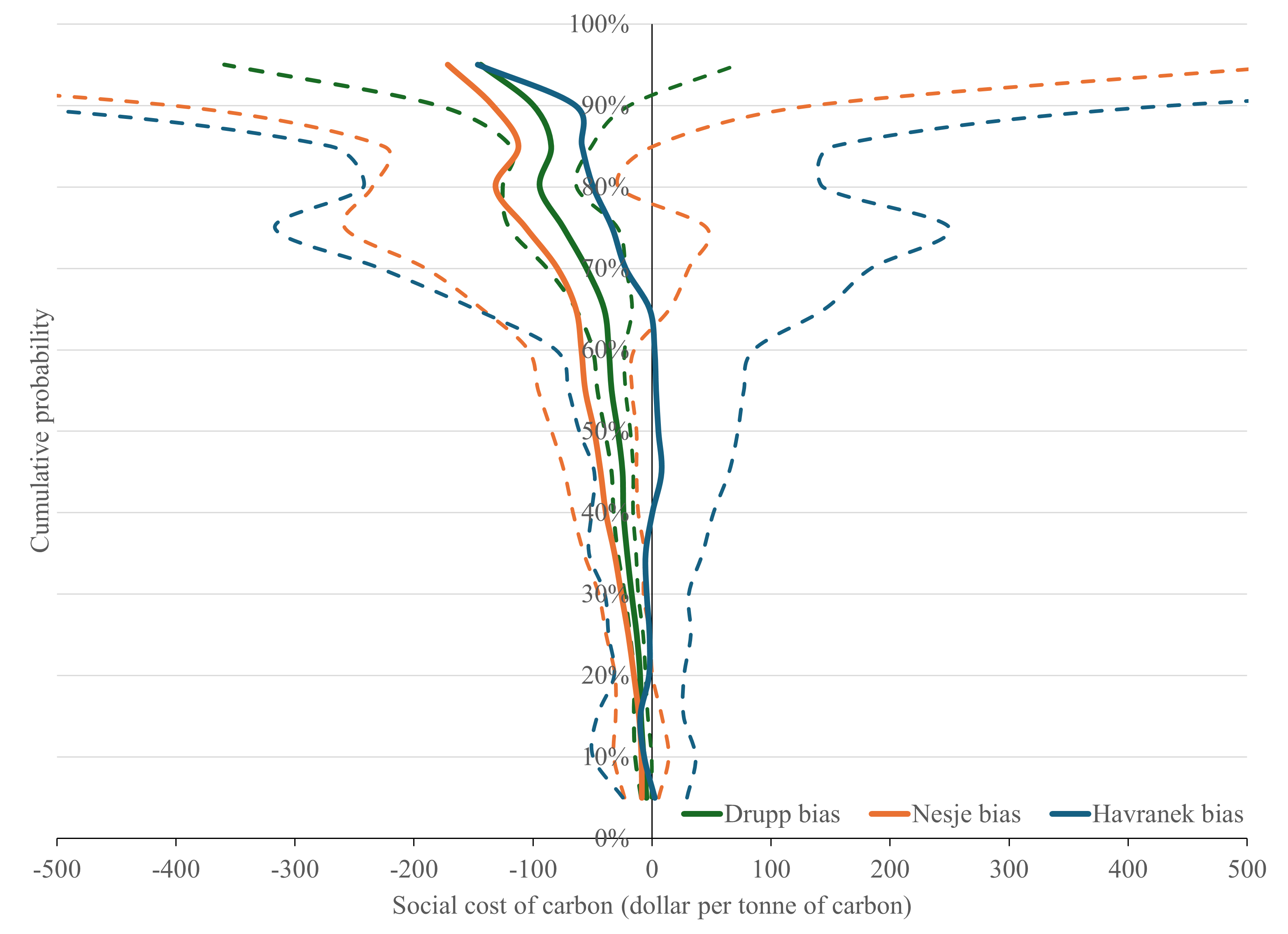}
    \caption*{\scriptsize The figures show the difference between the empirical cumulative distribution of the social cost of carbon and the emulated cumulative distribution with the recommendations on the pure rate of time preference and the inverse of the elasticity of intertemporal substitution of Drupp's experts and Nesje's experts and the revealed preferences in Havranek's and Matousek's meta-analyses. The dotted lines denote the 95\% confidence intervals.}
\end{figure}

\begin{figure}
    \centering
    \caption{The marginal impact of all three assumptions on the cumulative distribution of the social cost of carbon.}
    \label{fig:threeway}
    \includegraphics[width=\textwidth]{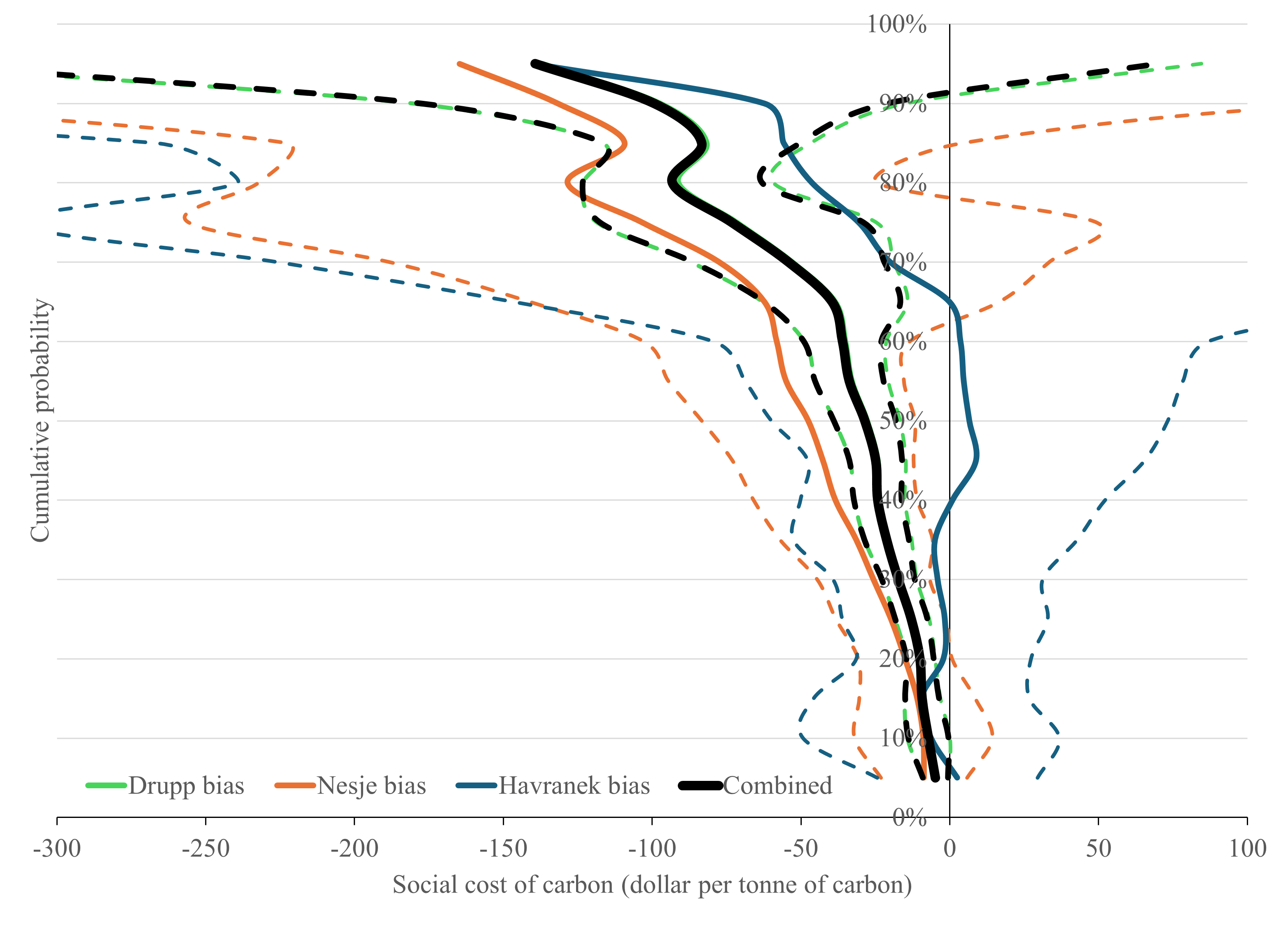}
    \caption*{\scriptsize The figures show the difference between the empirical cumulative distribution of the social cost of carbon and the emulated cumulative distribution with the recommendations on the pure rate of time preference and the inverse of the elasticity of intertemporal substitution of Drupp's experts and Nesje's experts and the revealed preferences in Havranek's and Matousek's meta-analyses, and the literature on impact of the economic impact of 2.5\celsius warming. The difference between the two is also shown with its 95\% confidence interval.}
\end{figure}

Adding the impact of climate change does not change the picture much; see Figure \ref{fig:threeway}. This is because the impact bias is relatively small (see Figure \ref{fig:emulimp}), as are its correlations with time and risk preferences.

Figure \ref{fig:threeway} further shows the combined bias.\footnote{Assuming normality, the variance is the harmonic mean variance $\sigma_c^{-2} = \sum_i \sigma_i^{-2}$ and the mean is the rescaled weighted mean $\mu_c = \sigma_c^2 \sum_i \mu_i \sigma_i^{-2}$.} The combined bias is very similar to the Drupp bias, as it has the narrowest confidence interval. Combining all assumptions implies that the literature \emph{under}estimates the social cost of carbon by \$5/tC (s.e. \$2/tC) at the 5\%ile, by \$29/tC (s.e. \$5/tC) at the 5\%ile, and by \$139/tC (s.e. \$105/tC) at the 95\%ile.

\section{Conclusion}
\label{sc:conclude}
Meta-emulation has three key advantages. First, it allows one to rapidly assess the implications of using alternative assumptions. Second, it tests whether the effect of said alternative assumptions is statistically significant. Third, it does so for the entire distribution of the social cost of carbon, rather than its central tendency.

Although meta-emulation is less demanding for the user than adopting and running an integrated assessment model, it is not trivial. This is demonstrated \href{https://richardtol.shinyapps.io/MetaSCC/}{here}. The \textsc{shiny} version of the emulator uses a coarsened probability distribution of the pure rate of time preference, but the user still needs to input 5 parameters. A fine-grained distribution for three parameters, including their correlations, may be too much to ask from most. However, a new survey or updated meta-analysis can readily be added and its impact on the social cost of carbon assessed.

Caveats and further research coalesce. Richer specifications can be tried, of course at the expense of the current linear and easy-to-interpret relationships. Any empirical analysis is limited by the availability of data. The two expert surveys used here are small and unrepresentative. One of the meta-analyses is old, the other restricted to a narrow set of studies. Studies of the social cost of carbon typically consider a few core parameters plus a set of assumptions that are almost unique to that paper. \citet{Moore2024PNAS} identify many more assumptions that affect the social cost of carbon. I agree that these are important, but I am not convinced that there are enough observations to confidently estimate their effect on the mean social cost of carbon, let alone its quantiles. A hierarchical model of the mean for many parameters and the quantiles for a few is beyond the current paper. 

These caveats notwithstanding, at the time of writing it appears that the literature on the social cost of carbon has made assumptions, particularly on the pure rate of time preference and the elasticity of intertemporal substitution, that are at odds with the literature and have led to downward bias in the reported social cost of carbon.

\section{Code and data}
Code and data are on \href{https://github.com/rtol/metascc}{GitHub}.

\bibliography{master}

\newpage\appendix
\setcounter{page}{1}
\renewcommand{\thepage}{A\arabic{page}}
\setcounter{table}{0}
\renewcommand{\thetable}{A\arabic{table}}
\setcounter{figure}{0}
\renewcommand{\thefigure}{A\arabic{figure}}
\setcounter{equation}{0}
\renewcommand{\theequation}{A\arabic{equation}}

\section{Growth impacts}
A relatively small number, 579 out of 14,152, of estimates of the social cost of carbon are based on the assumption that climate change would \emph{directly} affect the growth rate of the economy. Other estimates assume that climate change impacts the level of economic activity and \emph{indirectly} affects the growth rate \citep{Fankhauser2005}. The number of observations is too small for quantile regression. Weighted least squares reveals that the social cost of carbon falls by \$1271/tC (s.e. \$1188/tC) for every reduction in the growth rate by one percentage point for warming of 1\celsius, controlling for the year of publication, the pure rate of time preference, and the inverse of the elasticity of intertemporal substitution; see Table \ref{tab:growth}.

\begin{figure}
    \centering
    \caption{The empirical distributions of the economic growth impact as used in the literature on the social cost of carbon and found in the literature on the impact of weather shocks on economic growth.}
    \label{fig:growth}
    \includegraphics[width=\textwidth]{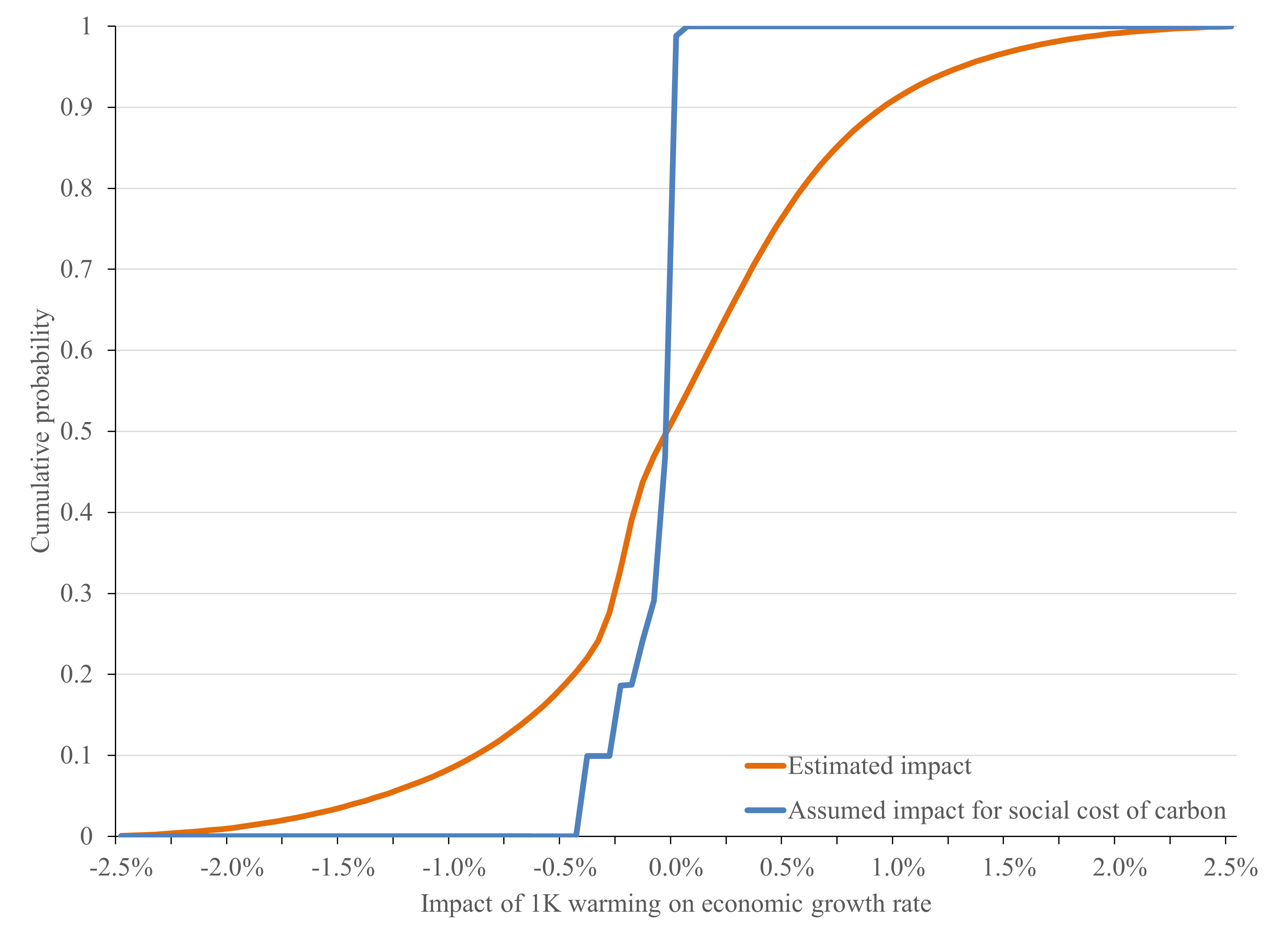}
\end{figure}

Figure \ref{fig:growth} shows the cumulative density function of the assumptions made in the social cost of carbon literature and contrasts this with the assumptions in the growth literature as surveyed by \citet{Tol2024EnPol}. Social cost estimates are based on a narrow set of studies and typically ignore parametric uncertainty. The range of estimates in the growth literature is vast. Correcting for this, the social cost of carbon \emph{over}estimates the social cost of carbon by \$146/tC (s.e. \$136/tC).

On average, controlling for time preferences and year of publication, estimates of the social cost of carbon are \$258/tC (s.e. \$139/tC) higher when based on growth rather than level impacts; see Table \ref{tab:growth}. This aligns with \citet[][Figure 5]{Moore2024PNAS}, who find a \$203/tC difference, with an interquartile range of \$51/tC to \$266/tC.\footnote{I am grateful to James Rising for providing the exact numbers.} However, as shown in Figure \ref{fig:growth}, more than half of the increase appears to be based on a selective reading of the empirical growth literature.

Few estimates of the social cost of carbon rely on the assumption that climate change affects the growth rate of the economy, and the assumed effect size is in a narrow range. These are the reasons why its coefficient is statistically insignificant in column (2) in Table \ref{tab:growth}. However, the corresponding dummy is significant in column (3). Columns (4) and (5) show that growth and level studies do not differ systematically on the assumed pure rate of time preference, but they do on the assumed elasticity of intertemporal substitution; growth studies are published later (column (6)). Table \ref{tab:growth} illustrates the dangers of random forests with small cell sizes and compositional effects.

\begin{table}[]
    \centering
    \caption{Regression results}
    \label{tab:coeffs}
    \scriptsize
    \begin{tabular}{l r r r r r r r r r r} \hline
         	&	mean	&	5\%	&	10\%	&	15\%	&	20\%	&	25\%	&	30\%	&	35\%	&	40\%	&	45\%	\\ \hline
PRTP	&	-206.00	&	-10.91	&	-16.54	&	-21.47	&	-23.14	&	-29.54	&	-39.29	&	-47.96	&	-54.92	&	-56.60	\\
	&	(21.33)	&	(3.098)	&	(4.956)	&	(4.174)	&	(3.377)	&	(3.972)	&	(4.054)	&	(5.489)	&	(5.891)	&	(6.516)	\\
EMUC	&	-139.20	&	-14.87	&	-7.20	&	-7.25	&	-20.02	&	-27.28	&	-33.66	&	-41.27	&	-56.93	&	-70.31	\\
	&	(36.11)	&	(5.543)	&	(8.868)	&	(7.468)	&	(6.042)	&	(7.106)	&	(7.254)	&	(9.822)	&	(10.54)	&	(11.66)	\\
Impact	&	-13.52	&	-1.26	&	-2.77	&	-3.66	&	-6.03	&	-5.70	&	-5.44	&	-5.20	&	-9.35	&	-12.50	\\
	&	(5.243)	&	(0.581)	&	(0.929)	&	(0.782)	&	(0.633)	&	(0.744)	&	(0.76)	&	(1.029)	&	(1.104)	&	(1.221)	\\
Year	&	4.58	&	0.98	&	1.49	&	1.60	&	2.14	&	2.43	&	2.31	&	1.94	&	2.30	&	3.04	\\
	&	(2.547)	&	(0.325)	&	(0.521)	&	(0.438)	&	(0.355)	&	(0.417)	&	(0.426)	&	(0.577)	&	(0.619)	&	(0.684)	\\
Constant	&	-8393	&	-1909	&	-2927	&	-3128	&	-4169	&	-4721	&	-4445	&	-3680	&	-4353	&	-5806	\\
	&	(5133)	&	(657.4)	&	(1052)	&	(885.7)	&	(716.6)	&	(842.8)	&	(860.3)	&	(1165)	&	(1250)	&	(1383)	\\ \hline
	&	median	&	55\%	&	60\%	&	65\%	&	70\%	&	75\%	&	80\%	&	85\%	&	90\%	&	95\%	\\ \hline
PRTP	&	-65.71	&	-77.48	&	-82.88	&	-91.82	&	-125.70	&	-170.00	&	-215.30	&	-193.40	&	-227.50	&	-328.80	\\
	&	(7.666)	&	(8.467)	&	(9.57)	&	(16.9)	&	(23.72)	&	(32.63)	&	(22.21)	&	(24.32)	&	(57.62)	&	(154.2)	\\
EMUC	&	-75.64	&	-84.77	&	-88.42	&	-91.30	&	-95.40	&	-123.70	&	-146.00	&	-109.40	&	-135.70	&	-115.70	\\
	&	(13.72)	&	(15.15)	&	(17.12)	&	(30.24)	&	(42.44)	&	(58.38)	&	(39.75)	&	(43.51)	&	(103.1)	&	(276)	\\
Impact	&	-19.62	&	-20.80	&	-23.41	&	-31.51	&	-38.07	&	-51.28	&	-55.60	&	-53.24	&	-40.36	&	-122.00	\\
	&	(1.437)	&	(1.587)	&	(1.794)	&	(3.167)	&	(4.446)	&	(6.116)	&	(4.164)	&	(4.558)	&	(10.8)	&	(28.91)	\\
Year	&	3.27	&	3.51	&	3.63	&	4.12	&	4.51	&	5.30	&	8.18	&	7.60	&	8.74	&	4.53	\\
	&	(0.805)	&	(0.889)	&	(1.005)	&	(1.775)	&	(2.491)	&	(3.427)	&	(2.333)	&	(2.554)	&	(6.053)	&	(16.2)	\\
Constant	&	-6243	&	-6685	&	-6901	&	-7842	&	-8524	&	-9942	&	-15535	&	-14399	&	-16414	&	-7523	\\
	&	(1627)	&	(1797)	&	(2031)	&	(3586)	&	(5033)	&	(6924)	&	(4714)	&	(5160)	&	(12228)	&	(32728)	\\  \hline
    \end{tabular}
    \caption*{\scriptsize Regression and quantile regression results. The dependent variable is the social cost of carbon. The explanatory variables are the pure rate of time preference, tne elasticity of the marginal utility of consumption, the impact of 2.5\celsius{} warming, and the year of publication. Numbers in brackets are the standard errors of the estimated coefficients.}
\end{table}

\begin{table}
\centering
\caption{Regression results: Level v growth impacts}
\label{tab:growth}
\footnotesize
\begin{tabular}{lcccccc} \hline
& SCC & SCC & SCC & PRTP & EIS & Year \\ \hline
PRTP & -206.0*** & -252.1 & -191.3*** &  &  &  \\
 & (21.33) & (280.5) & (20.55) &  &  &  \\
EIS & -139.2*** & 294.7 & -77.52*** &  &  &  \\
 & (36.11) & (370.5) & (28.68) &  &  &  \\
Year & 4.577* & 33.33 & 6.687*** & -0.0201*** & 0.000116 &  \\
 & (2.547) & (53.48) & (2.539) & (0.00383) & (0.00259) &  \\
Level & -13.52** &  &  &  &  &  \\
 & (5.243) &  &  &  &  &  \\
Growth &  & -1,271 &  &  &  &  \\
 &  & (1,188) &  &  &  &  \\
Level (dummy) &  &  & -44.69 & -0.148 & 0.170** & -4.817*** \\
 &  &  & (70.17) & (0.102) & (0.0724) & (0.527) \\
Growth (dummy) &  &  & 303.1** & 0.0429 & 0.0426 & 2.705** \\
 &  &  & (119.9) & (0.182) & (0.132) & (1.316) \\
Constant & -8,393 & -66,681 & -12,675** & 42.14*** & 0.914 & 2,018*** \\
 & (5,133) & (108,028) & (5,127) & (7.724) & (5.227) & (0.468) \\
 &  &  &  &  &  &  \\
Observations & 902 & 38 & 1,027 & 1,051 & 1,144 & 1,518 \\
R\textsuperscript{2} & 0.121 & 0.092 & 0.106 & 0.026 & 0.005 & 0.065 \\ \hline
\multicolumn{7}{c}{ Standard errors in parentheses} \\
\multicolumn{7}{c}{ *** p$<$0.01, ** p$<$0.05, * p$<$0.1} \\
\end{tabular}
\end{table}

\end{document}